# Area and Throughput Trade-Offs in the Design of Pipelined Discrete Wavelet Transform Architectures

Sandro V. Silva & Sergio Bampi
*Federal University of Rio Grande do Sul - Instituto de Informática*
*Porto Alegre, Brazil*
e-mail: <svsilva,bampi>@inf.ufrgs.br

## Abstract

*The JPEG2000 standard defines the discrete wavelet transform (DWT) as a linear space-to-frequency transform of the image domain in an irreversible compression. This irreversible discrete wavelet transform is implemented by FIR filter using 9/7 Daubechies coefficients or a lifting scheme of factorizated coefficients from 9/7 Daubechies coefficients.*

*This work investigates the tradeoffs between area, power and data throughput (or operating frequency) of several implementations of the Discrete Wavelet Transform using the lifting scheme in various pipeline designs. This paper shows the results of five different architectures synthesized and simulated in FPGAs. It concludes that the descriptions with pipelined operators provide the best area-power-operating frequency trade-off over non-pipelined operators descriptions. Those descriptions require around 40% more hardware to increase the maximum operating frequency up to 100% and reduce power consumption to less than 50%. Starting from behavioral HDL descriptions provide the best area-power-operating frequency trade-off, improving hardware cost and maximum operating frequency around 30% in comparison to structural descriptions for the same power requirement.*

## 1. Introduction

In images from real sceneries (photography), also called still tone images, adjacent pixels are almost always correlated, meaning that there are redundant data in the image coding [1]. Image storage and transmission is costly, and image compression is mandatory in embedded applications.

The image compression techniques operate reducing or eliminating the data redundancies of still tone images [1]. Image compression techniques can compress with or without loss of data of original image. The lossy compression can achieve higher ratio than lossless compression. The image compressor uses a linear transform to change the image domain from space to frequency, removing the correlation between pixels. An image compression technique efficiency depends on the amount of energy the linear transform can concentrate in few bands. So, in a lossy image compressor, after the linear transform the large amount of coefficients that are close to zero are eliminated by the quantizer block, and the quantized coefficients are entropy-coded for achieving high compression ratio. The loss of information eliminated by the quantizer operation is moderately compensated by the interpolation nature of the human visual system.

In the JPEG2000 standard [2] a lossy standard for image compression is defined using the discrete wavelet transform (DWT) for space-to-frequency transformation. This domain change achieves higher compression rates than DCT, since a large amount of coefficient images are close to zero.

The DWT can be designed by a sub-band multiresolution [1]. This sub-band multiresolution is made by recursive use of 9/7 taps FIR filters with Daubechies coefficients. By a lifting factorization operation [3], a simpler approach can be used, reducing the hardware complexity.

Several efficient architectures to implement a hardware one-dimensional discrete wavelet transform (1D-DWT) already exist. Implementations by filter banks are presented in [4], [5] and implementations by lifting scheme are presented in [6], [7].

To prototype the DWT to a FPGA, both behavioral or structural description can be used. A behavioral description can be mapped/synthesized through specific internal resources of FPGAs, which results in different net-lists. So, to implement an Intellectual Property (IP) Core or an Application Specific Integrated Circuit (ASIC), the resulting net-list may just contain macro structures that are specific to the FPGA vendor device mapping tool. To avoid this dependence on macro blocks, a structural design is definitely the choice.

In this work we implement five designs of a Discrete Wavelet Transform (DWT) by lifting scheme using behavioral and structural descriptions. We compare the tradeoffs between area cost, throughput (maximum frequency) and power requirement by synthesis and simulation in the Quartus II tools. All these implementations are synthesized to Altera APEX 20KE FPGA devices [8].

The rest of the paper is organized as follows. Section 2 explains the operation of the DWT. Section 3 describes the design of the lifting DWT architectures. In section 4 the results about area cost and operating frequencies are presented. Finally, section 5 presents a summary discussion and concluding remarks.

## 2. Discrete Wavelet Transform

The wavelet transform is a linear transform that can operate in direct or inverse form. The wavelet transform approximates a function by representing it as a linear combination of two sets of coefficients *g* and *h* constructed from functions derived respectively by a scaling function $\phi(t)$ and a mother wavelet function $\psi(t)$ [9].

The two-dimensional wavelet transform is computed by recursive application of one-dimensional wavelet transform. In a 1D-DWT each octave computes two sub-bands from one original band and each of this sub-bands has half the number of coefficients input without data loss. In a 2D-DWT each octave computes four sub-bands and each of these ones has a quarter number of coefficient input. Figure 1 shows the input and output of one octave of 2D-DWT.

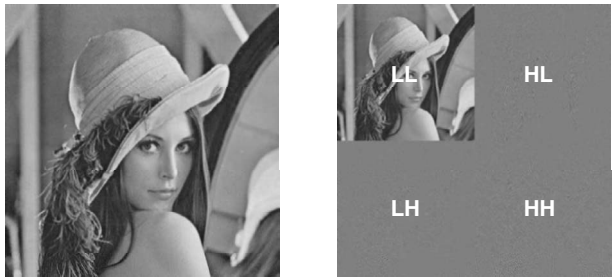

**Figure 1. One octave of 2D-DWT.**

The output coefficients from 1D-DWT are produced by application of two filters on data input samples, then producing two different output coefficient bands. A low-pass filter using h(x) coefficients producing an output band representing low-frequency data and a high-pass filter using g(x) coefficients producing an output band representing high frequency data.

The finite samples filtering present a problem of discontinuities its boundaries. So, the image boundary information could be lost if it were not treated properly. A simple method to eliminate this problem consists in mirroring the boundaries of the samples. The amount of samples mirroring depends on the depth of the low pass filter

The implementation of an irreversible DWT can be done by using the Daubechies biorthogonal wavelet coefficients [9], consisting of FIR filter coefficients for a 7-tap high-pass filter and a 9-tap low-pass filter, as shown in figure 2.

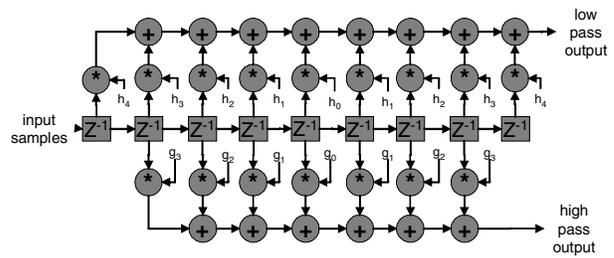

**Figure 2. DWT by 9/7 taps Daubechies FIR filter.**

The basic architecture defined in figure 2 requires 16 adders, 16 multipliers and 8 registers. This architecture has high area cost for a parallel implementation, thus several algorithms were developed to reduce this area cost.

The lifting DWT scheme presented in [3] has reduced computational complexity, so reducing the hardware cost to compute the DWT. This algorithm shown schematically in figure 3 was developed by a factorization of poliphase matrix from 9/7 Daubechies wavelet filter coefficients.

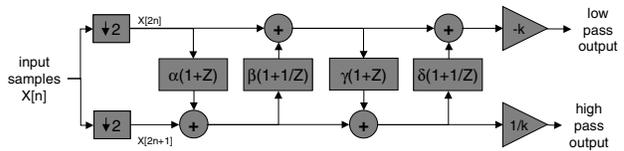

**Figure 3. Lifting DWT.**

## 3. Design of the architecture

The design of the 2D-DWT has three blocks: a 1D-DWT, memory and memory control blocks, as schematically shown in figure 4.

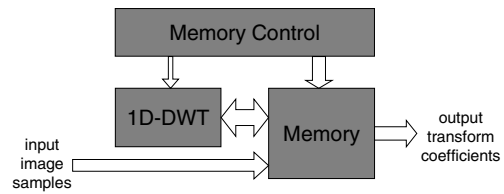

**Figure 4. 2D-DWT architecture.**

The input image samples are stored in memory, so the memory size needs to be as large as the image size. In the main step, the memory control addresses the coefficients of band to 1D-DWT and addresses the transformed coefficients back to the memory. After computation of all octaves, the transformed coefficients are transferred to next stage.

Generic 1D-DWT architecture can be designed as a pipelined structure following the lifting scheme. This basic design is shown in figure 5. It was implemented using five different approaches and synthesized to an Altera APEX 20KE FPGA device. This basic architecture can be designed with 6 multipliers, 8 adders and around 14 registers.

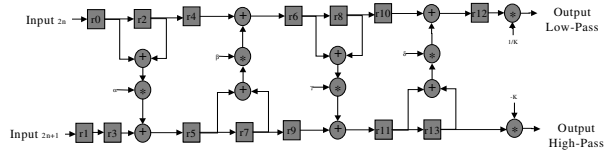

**Figure 5. Lifting 1D-DWT architecture.**

The multiplier constants used in the lifting scheme result from a factorization of coefficients from 9/7 Daubechies filter. The factorized coefficients in table 1 are represented as floating point values, integer ratios and fixed point binary in 2's complement.

**Table 1. Lifting coefficients constants.**

| Coefficient | Floating Point value | Integer rounded | Binary representation |
|---|---|---|---|
| alpha | -1.586 134 342 | -406/256 | 10.01101010 |
| beta | -0.052 980 118 | -14/256 | 11.11110010 |
| gamma | 0.882 911 075 | 226/256 | 00.11100010 |
| delta | 0.443 506 852 | 114/256 | 00.01110001 |
| -k | -1.230 174 105 | -314/256 | 10.11000101 |
| 1/k | 0.812 893 066 | 208/256 | 00.11010000 |

### 3.1 DWT architecture described with behavioral integer generic multipliers

The simplest approach of a Lifting 1D-DWT is described with integer generic multipliers. This architecture was described as an 8-stage integer pipeline with two data-flows. The first dataflow computes the even samples and the second dataflow computes the odd samples. The alpha and gamma multiplications are inserted in the flow, from even dataflow to odd dataflow, and beta and delta multiplications are inserted in the flow from odd dataflow to even dataflow, as shown in figure 3.

The implementation of an integer multiplier has lower cost than floating point multiplication. In these designs, all multiplication constants are integer divisions generated by rounding of floating point lifting constants. The result adjustment is done later by an 8-bit right shift. This rounding technique introduces a small error in coefficients, but this error is not significant to the transform result since all output coefficients are later to be quantized. The integer value coefficients used in these architectures are shown in table 1.

The error introduced by rounding can be measured by *peak signal to noise ratio* (PSNR) metric. The PSNR is calculated by quadratic differences between the original image pixels and the reconstructed image pixels, as showed in figure 6.

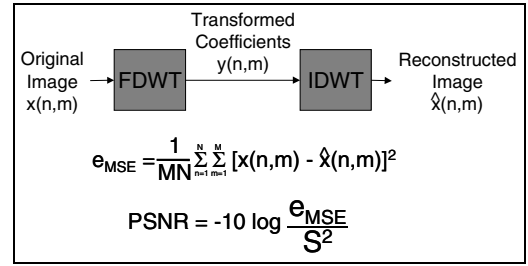

**Figure 6. PSNR calculation.**

From results of PSNR calculated in a tile of "Lena" image shown in table 2, the errors introduced in the output transform coefficients by the use of Lifting scheme with either floating-point coefficients or integer-rounded coefficients is always around 0.1dB. Hence, we can ignore these errors introduced by the rounding technique.

**Table 2. Measurement of rounding error.**

| Method | PSNR (dB) |
|---|---|
| FIR filter by floating point 9/7 Daubechies coefficients | 37.497 |
| FIR filter by integer rounded 9/7 Daubechies coefficients | 37.483 |
| Lifting scheme by floating point factorized coefficients | 37.094 |
| Lifting scheme by integer rounded factorized coefficients | 36.974 |

The bit length of DWT internal registers depends on their relative position inside the DWT pipeline.
- The registers located before alpha multiplication in the odd dataflow and before beta multiplication in the even dataflow store integers from -127 to 128 (signed 8-bits).
- The registers located after alpha multiplications and before gamma multiplication store integers from -530 to 530 (signed 11-bits).
- The registers located after beta multiplication and before delta multiplication store values from -184 to 184 (signed 9-bits).
- The registers located after gamma multiplication and before -k multiplication store values from -205 to 205 (signed 9-bits).
- The register located after delta multiplication and before division by k stores values from -366 to 366 (signed 10-bits).
- The register located at output data of the even dataflow (after division by k), corresponding to low

frequency of input image samples, stores values from -298 to 298 (signed 10-bits).
- The register located at output data of the odd dataflow (after multiplication by -k), corresponding to high frequency of input image samples, stores values from -252 to 252 (signed 9-bits). We kept these 9 bits, even though a low magnitude value is expected for this data output due to the nature of the transform of still-tone images.

### 3.2 DWT architecture described with behavioral shifted integer adders

Generic multipliers usually have high area cost. Multiplication by constant can be performed by shifted additions when the number of bits in the multiplication is large.

The decimal point showed in binary representation in Table 1 is for documenting the design of the control, as it is not considered in the hardware multiplier, which is fix-point and is later adjusted by 8-bit right shift.

By binary representation of multiplier constants presented in table 1, the multiplication by alpha needs 6 adders, the first one to perform r0+r2, the next four ones to perform the sum between second, fourth, sixth, seventh and two complement of tenth shifted partial products of r0+r2. The last one performs the sum with r3. Figure 6 shows how ordering the partial products are done to implement the multiplication by alpha.

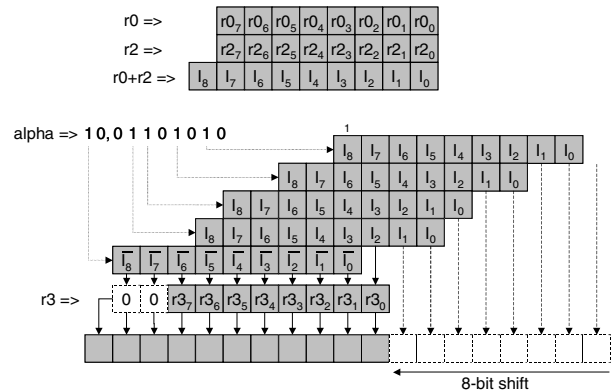

**Figure 7. Alpha multiplication by shifted adders.**

The multiplication by beta needed 8 adders, but one adder result can be re-used, reducing this stage to 7 adders. The multiplication by gamma needs 5 adders and the multiplication by delta needs 5 adders.

The -k equivalent constant has 5 high bits and this stage needed a simple multiplication, so 4 adders can perform the multiplication by -k. The 1/k equivalent has 3 high bits, so 2 adders can perform the multiplication by 1/k.

Hence, this design promotes only integer sums, reducing the area cost thus increasing the maximum operating frequency.

After all sums of each stage, the generated value is a real number, to generate the integer value on output stage a truncation promoted by an 8-bit right shift occur, as shown in figure 7.

### 3.3 DWT architecture described with behavioral pipeline of shifted integer adders

The architecture of lifting DWT can be naturally pipelined, but the add/shift stages represent the worst delay path between registers. Pipelining these stages increases the data throughput (operating frequency).

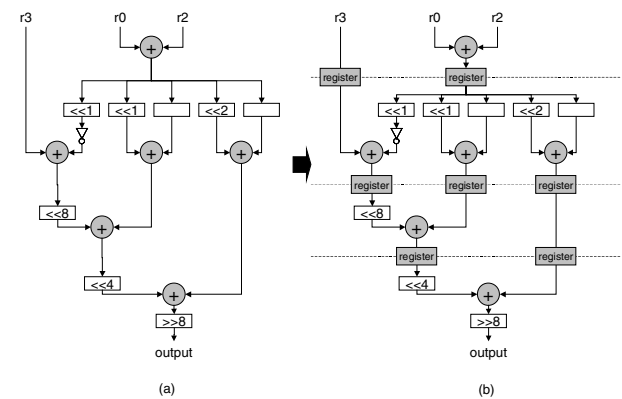

**Figure 8. Arithmetic stage structure of alpha multiplication.**

The original arithmetic stage structure to perform the multiplication by alpha is shown in figure 8(a). The pipelining of this stage is straightforward - as the insertion of some registers inside the logic shows in figure 8(b). There is just one sum operation at each pipeline stage.

Hence, pipelining increases the area cost of the architecture, but reduces the worst delay path between registers, increasing the throughput rate of the DWT. The computation of each stage will be faster than architectures presented in section 3.1 and 3.2 and the power consumption should be reduced, allowing the trade-offs we explore later.

### 3.4 DWT architecture described with structural shifted integer adders

A behavioral implementation is the simplest choice to improve a design to synthesize this architecture in a FPGA device, by using specific megacore functions to implement the multipliers and adders. But should it be required to synthesize this architecture in an ASIC, all structures must be defined structurally.

This implementation approach is similar to description in section 3.2 although the adders are structurally described. Except the alpha multiplications, the other multiplication blocks need to compute with signed operators, so the left bits from most significant bit (MSB) of an operator are replicated in the MSB.

All adders are implemented using Full Adder as basic blocks. The register adders are implemented by an n-bit set of full adder blocks. The length of each adder depends on the length of its inputs and length of the partial product shifted.

### 3.5 DWT architecture described with structural pipeline of shifted integer adders

This approach implementation has similarities with architecture described in section 3.3 and the structural description of adders described in section 3.4.

This architecture has 21 pipeline stages, as the design described in section 3.3, although the stages of sum and multiplications are designed with structural full adder. As the architecture of section 3.3, each complete sum operation is done at just one pipeline stage. With this improvement, the architecture achieves higher operating frequency at reduced power consumption.

The structural implementation of adders require more area cost than behavioral implementation, so it is expected that this implementation has higher area cost, although its throughput is higher than other architectures without pipelined arithmetic stages.

The main advantage of this architecture description is being independent of the prototyping hardware, i.e. this architecture can be prototyped onto any FPGA device or to an ASIC.

## 4 Implementation Results

Here we present all implementation results of 1D-DWT architectures presented in section 3. All VHDL descriptions we developed were compiled and simulated in an APEX20KE FPGA device from Altera. Table 3 shows the results of area cost, maximum data throughput (maximum operating frequency), number of pipeline stages and power consumption estimated at a given reference frequency (15MHz), set by the slowest architecture.

**Table 3. Implementation results.**

| Architecture | Area cost (LEs) | Maximum Operating frequency (MHz) | Power @15MHz (mW) | Pipeline stages |
|---|---|---|---|---|
| Design 1 | 781 | 16.6 | 310 | 8 |
| Design 2 | 480 | 44.0 | 248 | 8 |
| Design 3 | 766 | 157 | 105 | 21 |
| Design 4 | 701 | 54.4 | 232 | 8 |
| Design 5 | 1002 | 105 | 91.4 | 21 |

The design 1 is a behavioral implementation with integer multipliers. This is the simplest architecture to implement in VHDL. The architecture synthesized from the behavioral description incurred the highest power consumption simulates at 15MHz for comparison.

The smallest area cost architecture is the design 2, consisting a behavioral description using integer-shifted adders, as showed in figure 7. This reduced area cost is due to the fact that the synthesis utilizes fast carry chain propagation and the number of pipeline stages is limited to 8. So an 8-bit adder is mapped onto just 8 Logic Elements (LEs). In simulation of a tile of "Lena" at 40MHz, this architecture consumes 626mW, in other words, the 2.7 times increase in operating frequency over the first architecture provides a 2 times increase in its power consumption.

The design 3 is a behavioral implementation with pipelined integer shifted adders, resulting in a 21-stage pipeline. This is an architecture that has about 1.6 times the area cost of design 2 and has a maximum operating frequency that is 3 times larger. When running at 128MHz, this architecture consumes 808mW, representing 1.3 times the power consumption of the second architecture when this operates at less than one-third of the frequency (40MHz). So, the architecture of design 3 definitely shows a better area-power compromise per MHz than the second architecture.

The design 4 is a structural implementation with shifted integer adders. This architecture does not use the fast carry chain propagation, so an 8-bit adder requires 16 Logic Elements (LEs). It is expected the design 4 would have 2 times the area cost and same maximum operating frequency as design 2. But the result is the design 4 has 1.5 times the area cost and higher maximum frequency than design 2. The simulations at higher operation frequencies show that these two architectures maintain equivalent power consumption at same frequencies.

The design 5 is a structural implementation with pipelined shifted integer adders and it is an improvement of design 4. It presented the lowest power consumption of all 5 designs at the same frequency. As in design 3, design 5 has higher area cost, although it has higher maximum operating frequency. One could have expected both designs 3 and 5 to have about the same maximum frequency, which was not the case. This difference is because design 3 behavioral description can use the resources of fast carry chain propagation in the FPGA structure. The simulations show that this architecture has 15% less power consumption than design 3 at same operating frequency, requiring 476mW at 95MHz.

The architectures presented in [5], which is closest to ours, is implemented by filter banks using 785LEs at maximum operating frequency of 85.5 MHz. Comparing with our designs, the design 2 has half of area cost and its maximum operating frequency is nearly half, so they stand in different trade-off points. The design 3 we developed using deeper pipeline has the same area cost and its maximum operating frequency is double that of [5]. In this case, our architecture presents considerably larger throughput with the same area and possibly higher power. Power could not be compared against the work in the literature.

## 5. Conclusions

As presented in the results of table 2, the implementation of a lifting 1D-DWT using integer rounded factorized coefficients can be done without significant loss in image quality (less than 1dB of degradation in PSNR).

The implementation of a 1D-DWT architecture synthesized to a FPGA presents better results when using an initial behavioral description that takes advantage of the structural resources of FPGAs, such as the fast carry chain propagation.

The designs with pipelined operators (design 3 and design 5) have, in both cases - behavioral or structural - shown a better tradeoff between area cost and the throughput resulting from higher operating frequency. Moreover, the designs with pipelined operators reduced power consumption around 40%, at the expense of circa 40-60% more LEs (dependent on the type of description - 40% for the purely structural description and 60% for the behavioral description).

The structural HDL descriptions have worse trade-off area cost per maximum operating frequency than behavioral descriptions. The structural description presents around 30% more hardware and 30% less for the maximum operating frequency than behavioral description at same power requirement. This is because the technology mapping with commercial FPGA tools is optimized to map onto specific cores embedded in the particular FPGA family. To map the prototype to an 1D-DWT ASIC, the structural description is certainly a better starting point.

## Acknowledgements

The grant PDI-TI for CNPQ research agency is gratefully acknowledged.